\documentclass[twocolumn,preprintnumbers,amsmath,amssymb,superscriptaddress]{revtex4-1}
\usepackage{graphicx}
\usepackage{dcolumn}
\usepackage{bm}
\usepackage{natbib}

\def\apjs{Astrophys. J. Supp.}
\def\mnras{Mon.Not.Roy.As.Soc.}

\def\ltsima{$\; \buildrel < \over \sim \;$}
\def\simlt{\lower.5ex\hbox{\ltsima}}
\def\gtsima{$\; \buildrel > \over \sim \;$}
\def\simgt{\lower.5ex\hbox{\gtsima}}

\def\[{\begin{equation}}
\def\]{\end{equation}}
\def\m@th{\mathsurround=0pt }
\def\eqalign#1{\null\,\vcenter{\openup1\jot \m@th
 \ialign{\strut\hfil$\displaystyle{##}$&$\displaystyle{{}##}$\hfil
 \crcr#1\crcr}}\,}

\begin{document}
\title{The Scattering of Dark Matter and Dark Energy}

\date{\today}
\newcommand{\ud}{\mathrm{d}}
\newcommand{\fpe}{f_\perp}
\newcommand{\fpa}{f_\parallel}
\newcommand{\om}{\Omega_m}

\author{Fergus Simpson}
 \email{frgs@roe.ac.uk}
\affiliation{SUPA, Institute for Astronomy, University of
Edinburgh, Royal Observatory, Blackford Hill, Edinburgh EH9 3HJ}

\date{\today}

\begin{abstract}

We demonstrate how the two dominant constituents of the Universe, dark energy and dark matter, could possess a large scattering cross-section without considerably impacting observations.  Unlike interacting models which invoke energy exchange between the two fluids, the background cosmology remains unaltered, leaving fewer observational signatures.
Following a brief review of the scattering cross-sections between cosmologically significant particles, we explore the implications of an elastic interaction between dark matter and dark energy. The growth of large scale structure is suppressed, yet this effect is found to be weak due to the persistently low dark energy density. Thus we conclude that the dark matter-dark energy cross section may exceed the Thomson cross-section by several orders of magnitude.

\end{abstract}

\maketitle

\section{Introduction}

One of the most pressing issues in modern physics lies in the classification of dark energy. This is a phenomenon which not only appears to provide the bulk of the UniverseÕs energy content, but also gravitates in a repulsive manner, unlike any known substance. Prime candidates include the cosmological constant, scalar fields, and modifications to EinsteinÕs theory of gravity.

The first step in observationally distinguishing these models involves studying the cosmic geometry, since the cosmological constant makes a strong predictive statement on the trajectory of the cosmic expansion. Over the past decade, progress in this area has seen the redshift-distance relation tested by supernovae and Baryon Acoustic Oscillations (BAO) with a precision approaching $\sim 1 \%$. This translates into a bound on the dark energy equation of state $w \simeq -1 \pm 0.1$, where $w \equiv p/\rho$.

However, studying the expansion history alone is insufficient if we are to ever definitively exclude either scalar fields or modified theories of gravity. Therefore it is also of great importance to examine the growth of cosmic structure, an area which is attracting growing attention. This can be measured through various means such as redshift space distortions and weak gravitational lensing, though current constraints are relatively modest.

In performing this diagnosis of dark energy, we have implicitly been assuming that the physics within the dark sector of cosmology  - dark matter and dark energy - is purely gravitational. Yet what limitations can we place on their physical behaviour? While the precise nature of any microphysics is highly uncertain, the broader picture is one in which energy may be transferred either from dark energy to dark matter, or vice versa. Cosmologies with energy exchange have been extensively studied in the literature \cite{2000PhRvD..62d3511A, 2006PhRvD..73j3520B, valvi, valvi2, 2009PhRvD..79d3522C, jackson, sjp, 2001PhLB..521..133Z, baldi2010, LiBarrow}, and have been shown to leave characteristic signatures within observables such as the Integrated Sachs Wolfe effect, the Hubble constant $H_0$, and the growth of cosmic structure. Here we present a new class of models which don't leave as clear a cosmological signal as those which invoke energy exchange, since the comoving matter density remains fixed, and the background expansion is unaltered. Yet as we shall see, the growth of structure is readily suppressed by a drag term arising from elastic scattering between the dark matter and dark energy fluids.


\begin{figure*}
\includegraphics[width=140mm]{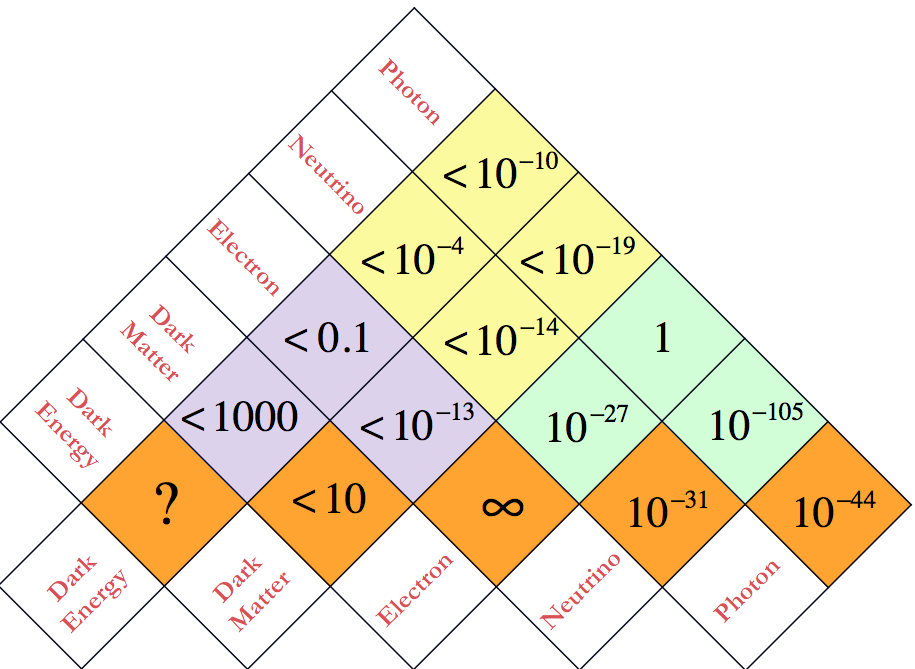}
\caption{A collection of cross-sections between cosmologically significant particles, in units of barns $(10^{-24} \rm{cm}^{2})$.  We assume a collisional energy associated with the era of recombination, $3000$K or equivalently $\sim 0.3$ eV.  The dark matter particle is taken to have a mass of  $10 \rm{~GeV}/\rm{c}^2$, and the dark energy equation of state $w=-0.9$.}  
\label{fig:pyramid}
\end{figure*}

\section{Cross Sections}

In general, an interaction between two particles may impart a transfer of momentum, a transfer of energy, or lead to the creation of new particles. Which of these might we expect to arise from the dark matter -- dark energy interaction?  Slow, low energy impacts (such as Thomson scattering and Rutherford scattering) often maintain elasticity, while relativistic velocities are more readily associated with inelastic events (such as Compton scattering and deep inelastic scattering). Given the extremely low dark energy density, and the nonrelativistic velocities associated with dark matter motions,  elastic scattering appears the simplest and most natural extension to dark sector physics. We need not restrict ourselves to a particular physical model of dark energy,  as the results obtained are largely independent of the microphysics involved in the scattering process.

The likelihood of scattering is quantified in terms of the cross section, which may be thought of as the effective target area as seen by an incident particle. We shall look to impose an upper bound on the scattering cross-section for dark matter -- dark energy interactions, and place this within the context of other cross-sections. Figure \ref{fig:pyramid} reviews the scattering cross-sections for a selection of cosmologically significant particles, which we briefly review in the subsections below. Many of these interactions exhibit a strong energy dependence, so in order to provide definitive values we adopt a cosmologically appropriate energy scale of $0.3 ~\rm{eV}$, corresponding to thermal collisions at the epoch of recombination $(z \sim 1100)$.  

One should also bear in mind that even with a fixed energy scale, interactions may exhibit a significant dependence on other factors such as spin or environment.  Thus for simplicity, and to facilitate a visual comparison, we focus on order of magnitude values.  

\subsection{Standard Model Scattering}

The values of cross-sections amongst Standard Model particles are generally well determined, albeit at much higher energy scales. The low energy values presented in Figure \ref{fig:pyramid} are either based on theoretical prediction, or a simple extrapolation from higher energies.

\begin{itemize} 
\item{At low energies, the {\bf photon-electron} interaction is governed by the Thomson cross section, $\sigma_T = 6.65 \times 10^{-25} \rm{~cm^2}$, which is of the order of one barn.}
\item{ {\bf Electron-electron} scattering is divergent due to the long range Coulomb interaciton.}
\item{ {\bf Photon-photon} scattering is strongly suppressed at energy scales below the electron rest mass, scaling as $E^6$.  This phenomenon has yet to be confirmed observationally, although recent constraints are approaching the required sensitivity \cite{PhysRevLett.96.083602}.}
\item{At such low energies, {\bf neutrino-neutrino} cross sections are poorly understood, here we provide a simple extrapolation from higher energy scales \cite{neutrino-neutrino}.} 
\item{ {\bf Neutrino-photon} scattering is of astrophysical importance, as it is capable of significantly influencing the evolution of stars and the dynamics of supernovae. However at sub-keV energy scales,  the elastic process dominates, leaving  $\sigma\left(\nu \gamma \rightarrow \nu \gamma \gamma \right) \ll \sigma \left( \nu \gamma \rightarrow \nu \gamma \right)$ \cite{neutrino-photon}} .
\item{Similarly, elastic scattering from neutral current interactions provides a prescription for the {\bf neutrino-electron} value \cite{ParticleReview}.}
\end{itemize}

\subsection{Dark - Standard Scattering}

No direct detection of either dark matter or dark energy has yet been made, so we are limited to applying upper bounds to these values. However interactions between the dark sector and standard model particles are quite restricted. 

For quoted bounds involving dark matter, these scale linearly as the particle mass, taken here to be $10 \rm{~GeV/c}^2$. 

\begin{itemize} 
\item{The {\bf dark matter-neutrino} bound is based on the detection of neutrinos arriving from SN1987A \cite{mangano}, which were not appreciably scattered by the intervening dark matter. Applying this analysis to the projected dark energy density leaves a significantly more modest constraint. }
\item{The Thomson optical depth established from observations of the Cosmic Microwave Background anisotropies places a strong lower bound on the mean free path of {\bf photons}. This acts as a limit on their interactions with the dark sector.}
\item{If {\bf electrons} were tightly coupled to dark matter or dark energy, this would impact on Cosmic Microwave Background anisotropies.}

\end{itemize}

\subsection{Dark Scattering}

We are left with just three components.

\begin{itemize} 
\item{The case of {\bf dark matter self-interactions} has been well studied \cite{ostriker-cross, spergel-steinhardt,mark-bullet}.  This upper bound stems from the disruption of subhalos which would occur near the centre of clusters. Note that bulk motions are unaffected, as only incoherent motions lead to scattering. This differs markedly from the case of dark matter-dark energy scattering, which we explore in detail in the following section. }
\item{In order to maintain stable density pertubations, dark energy is required to have some internal degrees of freedom. There may therefore be some lower bound on its self-interaction, but our extremely limited understanding of dark energy physics leaves {\bf dark energy - dark energy} scattering the most uncertain component of Figure \ref{fig:pyramid}.}
\item{Finally, the {\bf dark matter - dark energy} cross section, which provides the focus of this work. This weak bound is derived from the impact incurred on the growth of large scale structure, as outlined in the following section.}
\end{itemize}

\begin{figure}[t]
\includegraphics[width=80mm]{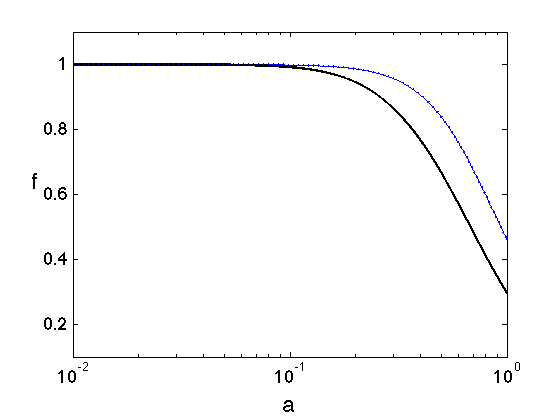}
\caption{The logarithmic growth rate of linear dark matter perturbations, when subject to elastic scattering with the dark energy fluid. For this configuration the particle mass $m_D = 10 \rm{~GeV/c^2}$, and $w=-0.9$. The solid line corresponds to a cross section of $\sigma_D=500~\rm{b}$, showing a suppression of growth at late times compared to the dotted line with no scattering $(\sigma_D=0)$.}
\label{fig:momentum}
\end{figure}

\section{Observational Impact} \label{sec:elastic}


As discussed in the previous section, indications from known physics suggest that \emph{elastic scattering} is the most abundant process at the energy scales of interest. Therefore we shall explore a scenario in which dark matter scatters elastically within the dark energy fluid. 

In order to define a cross section we must quantise dark energy in some manner. There are two regimes of interest - one in which the effective mass of dark energy is much greater or one in which it is much less than dark matter. For instance,  if a physical dark energy exists in a ``solid" configuration akin to a network of domain walls or cosmic strings \cite{1999PhRvD..60d3505B} each dark matter particle experiences a finite mean free path before being subject to a dissipationless recoil off the more massive structure.  We shall explore the ``light" regime, treating the dark energy fluid as being comprised of relativistic particles, and assume the characteristic negative pressure arises via their self-interaction. In this toy model, the particles merely act as a proxy for the energy density. However ultimately our analysis of the macroscopic behaviour and the conclusions drawn are largely independent of the microphysics involved. 

We begin by quantifying the impact dark scattering has on the growth of cosmic structure. 

\subsection{Large Scale Structure}

The coupled differential equations governing the linear density and velocity perturbations $\delta, ~ \theta,$ (see eg \cite{mabert, valvi}) are now modified, and we utilise the subscripts $Q$ and $c$ to denote the dark energy and dark matter fluids respectively. Provided the dark energy quanta are light and relativistic (non-relativistic particles would serve to increase the permitted cross section), the velocity perturbation exhibits a new drag term 

\[ \label{eq:eq1}
\theta'_Q = 2 H \theta_Q - a n_D \sigma_D \Delta \theta + k^2 \phi + k^2 \frac{\delta_Q}{1+w}   ,
\]

\noindent where $n_D$ is the proper number density of dark matter particles, $\sigma_D$ the scattering cross section between dark matter and dark energy, and we have defined the velocity contrast $\Delta \theta \equiv \theta_Q - \theta_c$. The combination $n_D \sigma_D \Delta \theta$ represents the fraction of the dark energy quanta which are subject to scattering per unit time. This is somewhat analogous to the Thomson scattering term which couples baryons and photons. Conservation of momentum leads to a similar term arising in the equivalent equation for dark matter, and this introduces a dependence on the dark energy equation of state.

\[\label{eq:eq2}
\theta'_c = - H \theta_c + \frac{\rho_Q}{\rho_c} (1+w) a n_D \sigma_D \Delta \theta + k^2 \phi  ,
\]

\noindent while the remaining perturbation equations are unchanged from their conventional form

\[\label{eq:eq3}
\eqalign{
\delta'_Q =& - \left[ \left(1+w\right) + 9\frac{H^2}{k^2} \left(1-w^2\right) \right] \theta_Q   \cr + & ~ 3 (1+w) \phi' - 3H(1-w) \delta_Q   ,
}
\]

\[\label{eq:eq4}
\delta'_c = - \theta_c + 3 \dot{\phi}  .
\]

\noindent The Poisson equation provides the source term

\[
k^2 \phi = 4 \pi G a^2 \sum_i \rho_i \delta_i  ,
\]

\noindent where we sum over the dark matter, dark energy, and baryons.
For our purposes the baryons are relatively inert, remaining unscattered by the dark energy fluid.  The dark energy sound speed is taken to be $c_{s}^2=1$.  We work on scales sufficiently below the horizon, $kH \ll 1$ such that the dark energy's large sound speed acts to maintain a high degree of homogeneity. 

Previous studies of coupled dark energy models characterise the energy-momentum transfer in terms of the 4-vector $Q^\mu$, and chose to align it with either the dark energy or dark matter rest frames \cite{valvi,valvi2, 2009PhRvD..79d3522C, jackson, sjp}. Here we have effectively rotated $Q^\mu$ to be spacelike, such that $Q^0=0$. Since the comoving matter density is conserved, the background $H(z)$ behaves no differently from that of the standard $w \rm{CDM}$ model.

\begin{figure}[t]
\includegraphics[width=80mm]{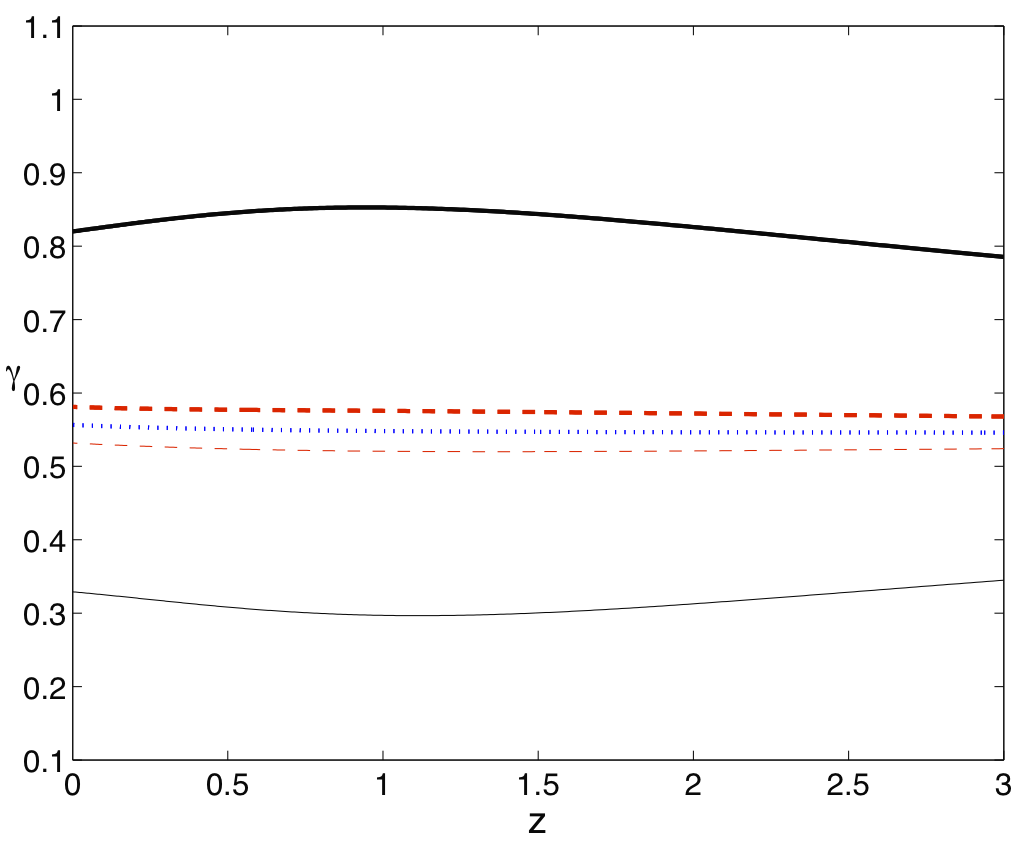}
\caption{The evolution in the growth index as a function of the scale factor. Thick solid and dashed lines correspond to models of dark energy with $w=-0.9$ and $w=-0.99$ respectively. As with Figure \ref{fig:momentum}, the dark matter - dark energy cross section is taken to be 500 b. The dotted line represents the standard case of zero scattering. Below the dotted line, the thin solid and dashed lines correspond to $w=-1.1$ and $w=-1.01$ models.}
\label{fig:gamma}
\end{figure}

\begin{figure}[t]
\includegraphics[width=80mm]{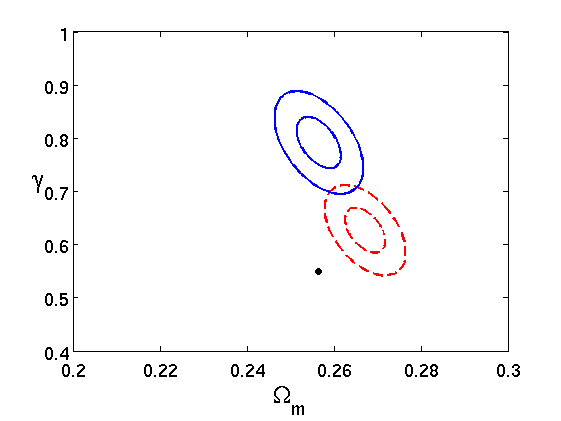}
\caption{The solid contours demonstrate the modification to the growth index induced by dark matter - dark energy scattering, with the cross section taken to be $\sigma_D=300$ b. The dashed contours provide an example of the bias which may be induced in the gravitational growth index $\gamma$ by the interacting model outlined in \cite{sjp}.  The standard model is indicated by the black dot. }
\label{fig:fig4}
\end{figure}

	The evolution of density perturbations in Figure \ref{fig:momentum} is evaluated by numerical integration of the six coupled differential equations, and is seen to depart significantly from the zero-scattering model. The anomalous behaviour in the growth rate $f \equiv \ud \ln \delta / \ud \ln a$  is more prevalent at late times, when there is simply more time available for interactions to occur. This fairly rapid onset of deceleration leads to the onset of baryon bias, with $\delta_b/\delta_c \simeq 1.1$ at low redshift. There are a number of potential tests for this baryon bias, from the composition of intra-cluster gas to the motions of tidally disrupted stellar streams. It has been noted that an apparent violation of the equivalence principle  of around $10\%$ is the upper bound based on current observations \cite{2006PhRvL..97m1303K}. 

The modified growth history may also be interpreted in terms of the growth index, defined such that

\[
\gamma \equiv \frac{\ud \ln f}{\ud \ln \Omega_m} \, .
\]

\noindent Ordinarily, General Relativity predicts $\gamma \simeq 6/11 - 15/2057 \,\Omega_\Lambda$ \cite{2010PhRvD..81j4020F}, yet in Figure \ref{fig:gamma} we see a significant departure from this value, due to scattering between dark matter and dark energy.  Provided $w>-1$, the drag term in (\ref{eq:eq2}) slows the growth of structure, enhancing the value of the growth index $\gamma$. If one considers $w<-1$, the sign of the drag term is reversed, thereby accelerating growth, however the physical interpretation of such a model is less clear.

\subsection{Redshift Space Distortions}

One of the leading techniques for studying the growth of large scale structure is redshift space distortions. The apparent anisotropy of the galaxy power spectrum provides a measure of the rate at which structure is forming on large scales.  In Figure \ref{fig:fig4} we demonstrate the rise in the growth index as measured by a galaxy survey at $z=0.5$, combined with Planck, following the Fisher matrix prescription outlined in \cite{simpsonp09}. This involves marginalising over the parameter set

\[
[w_0, w_a, \Omega_\Lambda,  \Omega_k, \Omega_m h^2, \Omega_b h^2, n_s, A_s, \beta, \gamma,
  \sigma_p].
\]

The standard cosmological parameters are taken to have fiducial values as
derived from WMAP5 \cite{2008arXiv0803.0547K}.
The contours plotted represent the estimated $1-$ and $2-\sigma$ likelihood contours.

 The solid line corresponds to a cross section $\sigma_D = 300 b$ and equation of state $w=-0.9$. Unlike energy exchange models, $\gamma$ is the only cosmological parameter subject to a bias. The dashed contours correspond to the energy exchange model outlined in \cite{sjp}, where dark energy decays into a form of dark matter. 

\subsection{Virialised Structures}

On smaller scales, consider a dark matter halo sat at rest in the dark energy frame. Elastic scattering acts in a similar manner to dark matter self-scattering, which would tend to isotropise  the halo. However, if we introduce a velocity-dependent cross-section, halos with a large peculiar velocity could exhibit an unusual behaviour.  Dark matter particles with motions aligned with the peculiar velocity would be subject to a greater retardation force,  and this may influence the orientation of the halo's ellipticity. A correlation of halo alignment with peculiar velocity may therefore be indicative of interactions in the dark sector.

\section{Discussion}

In the event that dark energy takes some physical form (neither a cosmological constant, nor a manifestation of new gravitational physics), then we might expect it to interact in some additional nongravitational manner. Of the interaction cross sections in Figure \ref{fig:pyramid} which are known, they are all \emph{non-zero} and predominantly \emph{elastic}. One might imagine that any such coupling within the dark sector must be extremely weak, in order to allow dark matter particles to experience a very long mean free path. However, owing to the persistently low energy density of dark energy, and the fairly low number density of dark matter particles, quite considerable cross sections are permitted. For an equation of state $w=-0.9$, this can exceed the Thomson cross section by two orders of magnitude before a significant impact is made on the growth of large scale structure. As we approach the limit $w=-1$, our constraint weakens further. 

Of course there are many subtleties which could alter the form of the interaction, such as a velocity-dependent cross-section. This model simply provides a demonstration of the vast volume of parameter space available for interactions between dark matter and dark energy to persist, and evade detection. 

In this class of models, we have identified a modification to the growth rate, and an induced baryon bias,  two features which are also associated with energy exchange. However, unlike models with energy exchange, the comoving matter density is conserved, and the expansion history remains unperturbed. In addition, a characteristic signature may reside in the alignment of dark matter halos with their direction of motion.

\section{Acknowledgements}
The author would like to thank Brendan Jackson, Andy Taylor, John Peacock, Pedro Ferreira, Constantinos Skordis, and Ed Copeland for helpful discussions, the anonymous referee for useful feedback, and acknowledges the support of an STFC rolling grant.

\bibliography{/Volumes/katrine.roe.ac.uk/Routines/dis}

\begin{thebibliography}{10}%
\makeatletter
\providecommand \@ifxundefined [1]{%
 \ifx #1\undefined \expandafter \@firstoftwo
 \else \expandafter \@secondoftwo
\fi
}%
\providecommand \@ifnum [1]{%
 \ifnum #1\expandafter \@firstoftwo
 \else \expandafter \@secondoftwo
\fi
}%
\providecommand \enquote [1]{``#1''}%
\providecommand \bibnamefont  [1]{#1}%
\providecommand \bibfnamefont [1]{#1}%
\providecommand \citenamefont [1]{#1}%
\providecommand\href[0]{\@sanitize\@href}%
\providecommand\@href[1]{\endgroup\@@startlink{#1}\endgroup\@@href}%
\providecommand\@@href[1]{#1\@@endlink}%
\providecommand \@sanitize [0]{\begingroup\catcode`\&12\catcode`\#12\relax}%
\@ifxundefined \pdfoutput {\@firstoftwo}{%
 \@ifnum{\z@=\pdfoutput}{\@firstoftwo}{\@secondoftwo}%
}{%
 \providecommand\@@startlink[1]{\leavevmode\special{html:<a href="#1">}}%
 \providecommand\@@endlink[0]{\special{html:</a>}}%
}{%
 \providecommand\@@startlink[1]{%
  \leavevmode
  \pdfstartlink
   attr{/Border[0 0 1 ]/H/I/C[0 1 1]}%
   user{/Subtype/Link/A<</Type/Action/S/URI/URI(#1)>>}%
  \relax
 }%
 \providecommand\@@endlink[0]{\pdfendlink}%
}%
\providecommand \url  [0]{\begingroup\@sanitize \@url }%
\providecommand \@url [1]{\endgroup\@href {#1}{\urlprefix}}%
\providecommand \urlprefix [0]{URL }%
\providecommand \Eprint[0]{\href }%
\@ifxundefined \urlstyle {%
  \providecommand \doi [1]{doi:\discretionary{}{}{}#1}%
}{%
  \providecommand \doi [0]{doi:\discretionary{}{}{}\begingroup
  \urlstyle{rm}\Url }%
}%
\providecommand \doibase [0]{http://dx.doi.org/}%
\providecommand \Doi[1]{\href{\doibase#1}}%
\providecommand \bibAnnote [3]{%
  \BibitemShut{#1}%
  \begin{quotation}\noindent
    \textsc{Key:}\ #2\\\textsc{Annotation:}\ #3%
  \end{quotation}%
}%
\providecommand \bibAnnoteFile [2]{%
  \IfFileExists{#2}{\bibAnnote {#1} {#2} {\input{#2}}}{}%
}%
\providecommand \typeout [0]{\immediate \write \m@ne }%
\providecommand \selectlanguage [0]{\@gobble}%
\providecommand \bibinfo [0]{\@secondoftwo}%
\providecommand \bibfield [0]{\@secondoftwo}%
\providecommand \translation [1]{[#1]}%
\providecommand \BibitemOpen[0]{}%
\providecommand \bibitemStop [0]{}%
\providecommand \bibitemNoStop [0]{.\EOS\space}%
\providecommand \EOS [0]{\spacefactor3000\relax}%
\providecommand \BibitemShut [1]{\csname bibitem#1\endcsname}%
\bibitem{2000PhRvD..62d3511A}%
  \BibitemOpen
  \bibfield{author}{%
  \bibinfo {author} {\bibfnamefont{L.}~\bibnamefont{{Amendola}}},\ }%
  \bibfield{journal}{%
  \bibinfo {journal} {\prd}\ }%
  \textbf{\bibinfo {volume} {62}},\ \bibinfo {pages} {043511} (\bibinfo {month}
  {Aug.}\ \bibinfo {year} {2000}),\
  \Eprint{http://arxiv.org/abs/astro-ph/9908023}{astro-ph/9908023}%
  \bibAnnoteFile{NoStop}{2000PhRvD..62d3511A}%
\bibitem{2006PhRvD..73j3520B}%
  \BibitemOpen
  \bibfield{author}{%
  \bibinfo {author} {\bibfnamefont{J.~D.}\ \bibnamefont{{Barrow}}}\ and\
  \bibinfo {author} {\bibfnamefont{T.}~\bibnamefont{{Clifton}}},\ }%
  \bibfield{journal}{%
  \Doi{10.1103/PhysRevD.73.103520}{\bibinfo {journal} {\prd}}\ }%
  \textbf{\bibinfo {volume} {73}},\ \bibinfo {pages} {103520} (\bibinfo {month}
  {May}\ \bibinfo {year} {2006}),\
  \Eprint{http://arxiv.org/abs/gr-qc/0604063}{gr-qc/0604063}%
  \bibAnnoteFile{NoStop}{2006PhRvD..73j3520B}%
\bibitem{valvi}%
  \BibitemOpen
  \bibfield{author}{%
  \bibinfo {author} {\bibfnamefont{J.}~\bibnamefont{{V{\"a}liviita}}}, \bibinfo
  {author} {\bibfnamefont{E.}~\bibnamefont{{Majerotto}}},\ and\ \bibinfo
  {author} {\bibfnamefont{R.}~\bibnamefont{{Maartens}}},\ }%
  \bibfield{journal}{%
  \Doi{10.1088/1475-7516/2008/07/020}{\bibinfo {journal} {Journal of Cosmology
  and Astro-Particle Physics}}\ }%
  \textbf{\bibinfo {volume} {7}},\ \bibinfo {pages} {20} (\bibinfo {month}
  {Jul.}\ \bibinfo {year} {2008}),\
  \Eprint{http://arxiv.org/abs/0804.0232}{arXiv:0804.0232}%
  \bibAnnoteFile{NoStop}{valvi}%
\bibitem{valvi2}%
  \BibitemOpen
  \bibfield{author}{%
  \bibinfo {author} {\bibfnamefont{J.}~\bibnamefont{{V{\"a}liviita}}}, \bibinfo
  {author} {\bibfnamefont{R.}~\bibnamefont{{Maartens}}},\ and\ \bibinfo
  {author} {\bibfnamefont{E.}~\bibnamefont{{Majerotto}}},\ }%
  \bibfield{journal}{%
  \Doi{10.1111/j.1365-2966.2009.16115.x}{\bibinfo {journal} {\mnras}}\ }%
  \textbf{\bibinfo {volume} {402}},\ \bibinfo {pages} {2355} (\bibinfo {month}
  {Mar.}\ \bibinfo {year} {2010}),\
  \Eprint{http://arxiv.org/abs/0907.4987}{arXiv:0907.4987}%
  \bibAnnoteFile{NoStop}{valvi2}%
\bibitem{2009PhRvD..79d3522C}%
  \BibitemOpen
  \bibfield{author}{%
  \bibinfo {author} {\bibfnamefont{S.}~\bibnamefont{{Chongchitnan}}},\ }%
  \bibfield{journal}{%
  \Doi{10.1103/PhysRevD.79.043522}{\bibinfo {journal} {\prd}}\ }%
  \textbf{\bibinfo {volume} {79}},\ \bibinfo {pages} {043522} (\bibinfo {month}
  {Feb.}\ \bibinfo {year} {2009}),\
  \Eprint{http://arxiv.org/abs/0810.5411}{arXiv:0810.5411}%
  \bibAnnoteFile{NoStop}{2009PhRvD..79d3522C}%
\bibitem{jackson}%
  \BibitemOpen
  \bibfield{author}{%
  \bibinfo {author} {\bibfnamefont{B.~M.}\ \bibnamefont{{Jackson}}}, \bibinfo
  {author} {\bibfnamefont{A.}~\bibnamefont{{Taylor}}},\ and\ \bibinfo {author}
  {\bibfnamefont{A.}~\bibnamefont{{Berera}}},\ }%
  \bibfield{journal}{%
  \Doi{10.1103/PhysRevD.79.043526}{\bibinfo {journal} {\prd}}\ }%
  \textbf{\bibinfo {volume} {79}},\ \bibinfo {pages} {043526} (\bibinfo {month}
  {Feb.}\ \bibinfo {year} {2009}),\
  \Eprint{http://arxiv.org/abs/0901.3272}{arXiv:0901.3272}%
  \bibAnnoteFile{NoStop}{jackson}%
\bibitem{sjp}%
  \BibitemOpen
  \bibfield{author}{%
  \bibinfo {author} {\bibfnamefont{F.}~\bibnamefont{{Simpson}}}, \bibinfo
  {author} {\bibfnamefont{B.~M.}\ \bibnamefont{{Jackson}}},\ and\ \bibinfo
  {author} {\bibfnamefont{J.~A.}\ \bibnamefont{{Peacock}}},\ }%
  \bibfield{journal}{%
  \bibinfo {journal} {ArXiv e-prints}}%
   (\bibinfo {month} {Apr.}\ \bibinfo {year} {2010}),\
  \Eprint{http://arxiv.org/abs/1004.1920}{arXiv:1004.1920}%
  \bibAnnoteFile{NoStop}{sjp}%
\bibitem{2001PhLB..521..133Z}%
  \BibitemOpen
  \bibfield{author}{%
  \bibinfo {author} {\bibfnamefont{W.}~\bibnamefont{{Zimdahl}}}, \bibinfo
  {author} {\bibfnamefont{D.}~\bibnamefont{{Pav{\'o}n}}},\ and\ \bibinfo
  {author} {\bibfnamefont{L.~P.}\ \bibnamefont{{Chimento}}},\ }%
  \bibfield{journal}{%
  \Doi{10.1016/S0370-2693(01)01174-1}{\bibinfo {journal} {Physics Letters B}}\
  }%
  \textbf{\bibinfo {volume} {521}},\ \bibinfo {pages} {133} (\bibinfo {month}
  {Nov.}\ \bibinfo {year} {2001}),\
  \Eprint{http://arxiv.org/abs/arXiv:astro-ph/0105479}{arXiv:astro-ph/0105479}%
  \bibAnnoteFile{NoStop}{2001PhLB..521..133Z}%
\bibitem{baldi2010}%
  \BibitemOpen
  \bibfield{author}{%
  \bibinfo {author} {\bibfnamefont{M.}~\bibnamefont{{Baldi}}},\ }%
  \bibfield{journal}{%
  \bibinfo {journal} {ArXiv e-prints}}%
   (\bibinfo {month} {May}\ \bibinfo {year} {2010}),\
  \Eprint{http://arxiv.org/abs/1005.2188}{arXiv:1005.2188}%
  \bibAnnoteFile{NoStop}{baldi2010}%
\bibitem{LiBarrow}%
  \BibitemOpen
  \bibfield{author}{%
  \bibinfo {author} {\bibfnamefont{B.}~\bibnamefont{{Li}}}\ and\ \bibinfo
  {author} {\bibfnamefont{J.~D.}\ \bibnamefont{{Barrow}}},\ }%
  \bibfield{journal}{%
  \bibinfo {journal} {ArXiv e-prints}}%
   (\bibinfo {month} {May}\ \bibinfo {year} {2010}),\
  \Eprint{http://arxiv.org/abs/1005.4231}{arXiv:1005.4231}%
  \bibAnnoteFile{NoStop}{LiBarrow}%
\bibitem{PhysRevLett.96.083602}%
  \BibitemOpen
  \bibfield{author}{%
  \bibinfo {author} {\bibfnamefont{E.}~\bibnamefont{Lundstr\"om}}, \bibinfo
  {author} {\bibfnamefont{G.}~\bibnamefont{Brodin}}, \bibinfo {author}
  {\bibfnamefont{J.}~\bibnamefont{Lundin}}, \bibinfo {author}
  {\bibfnamefont{M.}~\bibnamefont{Marklund}}, \bibinfo {author}
  {\bibfnamefont{R.}~\bibnamefont{Bingham}}, \bibinfo {author}
  {\bibfnamefont{J.}~\bibnamefont{Collier}}, \bibinfo {author}
  {\bibfnamefont{J.~T.}\ \bibnamefont{Mendon\ifmmode~\mbox{\c{c}}\else
  \c{c}\fi{}a}},\ and\ \bibinfo {author}
  {\bibfnamefont{P.}~\bibnamefont{Norreys}},\ }%
  \bibfield{journal}{%
  \Doi{10.1103/PhysRevLett.96.083602}{\bibinfo {journal} {Phys. Rev. Lett.}}\
  }%
  \textbf{\bibinfo {volume} {96}},\ \bibinfo {pages} {083602} (\bibinfo {month}
  {Mar}\ \bibinfo {year} {2006})%
  \bibAnnoteFile{NoStop}{PhysRevLett.96.083602}%
\bibitem{neutrino-neutrino}%
  \BibitemOpen
  \bibfield{author}{%
  \bibinfo {author} {\bibfnamefont{D.~F.}\ \bibnamefont{Physik}}, \bibinfo
  {author} {\bibfnamefont{B.}~\bibnamefont{Eberle}}, \bibinfo {author}
  {\bibfnamefont{P.}~\bibnamefont{Dr}},\ and\ \bibinfo {author}
  {\bibfnamefont{J.}~\bibnamefont{Bartels}}}%
   (\bibinfo {year} {2009}),\
  \url{http://citeseerx.ist.psu.edu/viewdoc/summary?doi=?doi=10.1.1.138.359}%
  \bibAnnoteFile{NoStop}{neutrino-neutrino}%
\bibitem{neutrino-photon}%
  \BibitemOpen
  \bibfield{author}{%
  \bibinfo {author} {\bibfnamefont{D.~A.}\ \bibnamefont{{Dicus}}}\ and\
  \bibinfo {author} {\bibfnamefont{W.~W.}\ \bibnamefont{{Repko}}},\ }%
  \bibfield{journal}{%
  \Doi{10.1103/PhysRevLett.79.569}{\bibinfo {journal} {Physical Review
  Letters}}\ }%
  \textbf{\bibinfo {volume} {79}},\ \bibinfo {pages} {569} (\bibinfo {month}
  {Jul.}\ \bibinfo {year} {1997}),\
  \Eprint{http://arxiv.org/abs/arXiv:hep-ph/9703210}{arXiv:hep-ph/9703210}%
  \bibAnnoteFile{NoStop}{neutrino-photon}%
\bibitem{ParticleReview}%
  \BibitemOpen
  \bibfield{author}{%
  \bibinfo {author} {\bibfnamefont{K.}~\bibnamefont{Hagiwara}}, \bibinfo
  {author} {\bibfnamefont{K.}~\bibnamefont{Hikasa}}, \bibinfo {author}
  {\bibfnamefont{K.}~\bibnamefont{Nakamura}}, \bibinfo {author}
  {\bibfnamefont{M.}~\bibnamefont{Tanabashi}}, \bibinfo {author}
  {\bibfnamefont{M.}~\bibnamefont{Aguilar-Benitez}}, \bibinfo {author}
  {\bibfnamefont{C.}~\bibnamefont{Amsler}}, \bibinfo {author}
  {\bibfnamefont{R.~M.}\ \bibnamefont{Barnett}}, \bibinfo {author}
  {\bibfnamefont{P.~R.}\ \bibnamefont{Burchat}}, \bibinfo {author}
  {\bibfnamefont{C.~D.}\ \bibnamefont{Carone}}, \bibinfo {author}
  {\bibfnamefont{C.}~\bibnamefont{Caso}}, \bibinfo {author}
  {\bibfnamefont{G.}~\bibnamefont{Conforto}}, \bibinfo {author}
  {\bibfnamefont{O.}~\bibnamefont{Dahl}}, \bibinfo {author}
  {\bibfnamefont{M.}~\bibnamefont{Doser}}, \bibinfo {author}
  {\bibfnamefont{S.}~\bibnamefont{Eidelman}}, \bibinfo {author}
  {\bibfnamefont{J.~L.}\ \bibnamefont{Feng}}, \bibinfo {author}
  {\bibfnamefont{L.}~\bibnamefont{Gibbons}}, \bibinfo {author}
  {\bibfnamefont{M.}~\bibnamefont{Goodman}}, \bibinfo {author}
  {\bibfnamefont{C.}~\bibnamefont{Grab}}, \bibinfo {author}
  {\bibfnamefont{D.~E.}\ \bibnamefont{Groom}}, \bibinfo {author}
  {\bibfnamefont{A.}~\bibnamefont{Gurtu}}, \bibinfo {author}
  {\bibfnamefont{K.~G.}\ \bibnamefont{Hayes}}, \bibinfo {author}
  {\bibfnamefont{J.~J.}\ \bibnamefont{Herna`ndez-Rey}}, \bibinfo {author}
  {\bibfnamefont{K.}~\bibnamefont{Honscheid}}, \bibinfo {author}
  {\bibfnamefont{C.}~\bibnamefont{Kolda}}, \bibinfo {author}
  {\bibfnamefont{M.~L.}\ \bibnamefont{Mangano}}, \bibinfo {author}
  {\bibfnamefont{D.~M.}\ \bibnamefont{Manley}},\ and\ \bibinfo {author}
  {\bibfnamefont{A.~V.}\ \bibnamefont{Manohar}},\ }%
  \bibfield{journal}{%
  \Doi{10.1103/PhysRevD.66.010001}{\bibinfo {journal} {Phys. Rev. D}}\ }%
  \textbf{\bibinfo {volume} {66}},\ \bibinfo {pages} {010001} (\bibinfo {month}
  {Jul}\ \bibinfo {year} {2002})%
  \bibAnnoteFile{NoStop}{ParticleReview}%
\bibitem{mangano}%
  \BibitemOpen
  \bibfield{author}{%
  \bibinfo {author} {\bibfnamefont{G.}~\bibnamefont{{Mangano}}},\ }%
  \bibfield{journal}{%
  \Doi{10.1016/j.nuclphysbps.2007.02.026}{\bibinfo {journal} {Nuclear Physics B
  Proceedings Supplements}}\ }%
  \textbf{\bibinfo {volume} {168}},\ \bibinfo {pages} {34} (\bibinfo {month}
  {Jun.}\ \bibinfo {year} {2007}),\
  \Eprint{http://arxiv.org/abs/arXiv:astro-ph/0611887}{arXiv:astro-ph/0611887}%
  \bibAnnoteFile{NoStop}{mangano}%
\bibitem{ostriker-cross}%
  \BibitemOpen
  \bibfield{author}{%
  \bibinfo {author} {\bibfnamefont{O.~Y.}\ \bibnamefont{{Gnedin}}}\ and\
  \bibinfo {author} {\bibfnamefont{J.~P.}\ \bibnamefont{{Ostriker}}},\ }%
  \bibfield{journal}{%
  \Doi{10.1086/323211}{\bibinfo {journal} {\apj}}\ }%
  \textbf{\bibinfo {volume} {561}},\ \bibinfo {pages} {61} (\bibinfo {month}
  {Nov.}\ \bibinfo {year} {2001}),\
  \Eprint{http://arxiv.org/abs/arXiv:astro-ph/0010436}{arXiv:astro-ph/0010436}%
  \bibAnnoteFile{NoStop}{ostriker-cross}%
\bibitem{spergel-steinhardt}%
  \BibitemOpen
  \bibfield{author}{%
  \bibinfo {author} {\bibfnamefont{D.~N.}\ \bibnamefont{{Spergel}}}\ and\
  \bibinfo {author} {\bibfnamefont{P.~J.}\ \bibnamefont{{Steinhardt}}},\ }%
  \bibfield{journal}{%
  \Doi{10.1103/PhysRevLett.84.3760}{\bibinfo {journal} {Physical Review
  Letters}}\ }%
  \textbf{\bibinfo {volume} {84}},\ \bibinfo {pages} {3760} (\bibinfo {month}
  {Apr.}\ \bibinfo {year} {2000}),\
  \Eprint{http://arxiv.org/abs/arXiv:astro-ph/9909386}{arXiv:astro-ph/9909386}%
  \bibAnnoteFile{NoStop}{spergel-steinhardt}%
\bibitem{mark-bullet}%
  \BibitemOpen
  \bibfield{author}{%
  \bibinfo {author} {\bibfnamefont{M.}~\bibnamefont{{Markevitch}}}, \bibinfo
  {author} {\bibfnamefont{A.~H.}\ \bibnamefont{{Gonzalez}}}, \bibinfo {author}
  {\bibfnamefont{D.}~\bibnamefont{{Clowe}}}, \bibinfo {author}
  {\bibfnamefont{A.}~\bibnamefont{{Vikhlinin}}}, \bibinfo {author}
  {\bibfnamefont{W.}~\bibnamefont{{Forman}}}, \bibinfo {author}
  {\bibfnamefont{C.}~\bibnamefont{{Jones}}}, \bibinfo {author}
  {\bibfnamefont{S.}~\bibnamefont{{Murray}}},\ and\ \bibinfo {author}
  {\bibfnamefont{W.}~\bibnamefont{{Tucker}}},\ }%
  \bibfield{journal}{%
  \Doi{10.1086/383178}{\bibinfo {journal} {\apj}}\ }%
  \textbf{\bibinfo {volume} {606}},\ \bibinfo {pages} {819} (\bibinfo {month}
  {May}\ \bibinfo {year} {2004}),\
  \Eprint{http://arxiv.org/abs/arXiv:astro-ph/0309303}{arXiv:astro-ph/0309303}%
  \bibAnnoteFile{NoStop}{mark-bullet}%
\bibitem{1999PhRvD..60d3505B}%
  \BibitemOpen
  \bibfield{author}{%
  \bibinfo {author} {\bibfnamefont{M.}~\bibnamefont{{Bucher}}}\ and\ \bibinfo
  {author} {\bibfnamefont{D.}~\bibnamefont{{Spergel}}},\ }%
  \bibfield{journal}{%
  \Doi{10.1103/PhysRevD.60.043505}{\bibinfo {journal} {\prd}}\ }%
  \textbf{\bibinfo {volume} {60}},\ \bibinfo {pages} {043505} (\bibinfo {month}
  {Aug.}\ \bibinfo {year} {1999}),\
  \Eprint{http://arxiv.org/abs/arXiv:astro-ph/9812022}{arXiv:astro-ph/9812022}%
  \bibAnnoteFile{NoStop}{1999PhRvD..60d3505B}%
\bibitem{mabert}%
  \BibitemOpen
  \bibfield{author}{%
  \bibinfo {author} {\bibfnamefont{C.}~\bibnamefont{{Ma}}}\ and\ \bibinfo
  {author} {\bibfnamefont{E.}~\bibnamefont{{Bertschinger}}},\ }%
  \bibfield{journal}{%
  \Doi{10.1086/176550}{\bibinfo {journal} {\apj}}\ }%
  \textbf{\bibinfo {volume} {455}},\ \bibinfo {pages} {7} (\bibinfo {month}
  {Dec.}\ \bibinfo {year} {1995}),\
  \Eprint{http://arxiv.org/abs/arXiv:astro-ph/9506072}{arXiv:astro-ph/9506072}%
  \bibAnnoteFile{NoStop}{mabert}%
\bibitem{2006PhRvL..97m1303K}%
  \BibitemOpen
  \bibfield{author}{%
  \bibinfo {author} {\bibfnamefont{M.}~\bibnamefont{{Kesden}}}\ and\ \bibinfo
  {author} {\bibfnamefont{M.}~\bibnamefont{{Kamionkowski}}},\ }%
  \bibfield{journal}{%
  \Doi{10.1103/PhysRevLett.97.131303}{\bibinfo {journal} {Physical Review
  Letters}}\ }%
  \textbf{\bibinfo {volume} {97}},\ \bibinfo {pages} {131303} (\bibinfo {month}
  {Sep.}\ \bibinfo {year} {2006}),\
  \Eprint{http://arxiv.org/abs/arXiv:astro-ph/0606566}{arXiv:astro-ph/0606566}%
  \bibAnnoteFile{NoStop}{2006PhRvL..97m1303K}%
\bibitem{2010PhRvD..81j4020F}%
  \BibitemOpen
  \bibfield{author}{%
  \bibinfo {author} {\bibfnamefont{P.~G.}\ \bibnamefont{{Ferreira}}}\ and\
  \bibinfo {author} {\bibfnamefont{C.}~\bibnamefont{{Skordis}}},\ }%
  \bibfield{journal}{%
  \Doi{10.1103/PhysRevD.81.104020}{\bibinfo {journal} {\prd}}\ }%
  \textbf{\bibinfo {volume} {81}},\ \bibinfo {pages} {104020} (\bibinfo {month}
  {May}\ \bibinfo {year} {2010}),\
  \Eprint{http://arxiv.org/abs/1003.4231}{arXiv:1003.4231}%
  \bibAnnoteFile{NoStop}{2010PhRvD..81j4020F}%
\bibitem{simpsonp09}%
  \BibitemOpen
  \bibfield{author}{%
  \bibinfo {author} {\bibfnamefont{F.}~\bibnamefont{{Simpson}}}\ and\ \bibinfo
  {author} {\bibfnamefont{J.~A.}\ \bibnamefont{{Peacock}}},\ }%
  \bibfield{journal}{%
  \Doi{10.1103/PhysRevD.81.043512}{\bibinfo {journal} {\prd}}\ }%
  \textbf{\bibinfo {volume} {81}},\ \bibinfo {pages} {043512} (\bibinfo {month}
  {Feb.}\ \bibinfo {year} {2010}),\
  \Eprint{http://arxiv.org/abs/0910.3834}{arXiv:0910.3834}%
  \bibAnnoteFile{NoStop}{simpsonp09}%
\bibitem{2008arXiv0803.0547K}%
  \BibitemOpen
  \bibfield{author}{%
  \bibinfo {author} {\bibfnamefont{E.}~\bibnamefont{{Komatsu}}}, \bibinfo
  {author} {\bibfnamefont{J.}~\bibnamefont{{Dunkley}}}, \bibinfo {author}
  {\bibfnamefont{M.~R.}\ \bibnamefont{{Nolta}}}, \bibinfo {author}
  {\bibfnamefont{C.~L.}\ \bibnamefont{{Bennett}}}, \bibinfo {author}
  {\bibfnamefont{B.}~\bibnamefont{{Gold}}}, \bibinfo {author}
  {\bibfnamefont{G.}~\bibnamefont{{Hinshaw}}}, \bibinfo {author}
  {\bibfnamefont{N.}~\bibnamefont{{Jarosik}}}, \bibinfo {author}
  {\bibfnamefont{D.}~\bibnamefont{{Larson}}}, \bibinfo {author}
  {\bibfnamefont{M.}~\bibnamefont{{Limon}}}, \bibinfo {author}
  {\bibfnamefont{L.}~\bibnamefont{{Page}}}, \bibinfo {author}
  {\bibfnamefont{D.~N.}\ \bibnamefont{{Spergel}}}, \bibinfo {author}
  {\bibfnamefont{M.}~\bibnamefont{{Halpern}}}, \bibinfo {author}
  {\bibfnamefont{R.~S.}\ \bibnamefont{{Hill}}}, \bibinfo {author}
  {\bibfnamefont{A.}~\bibnamefont{{Kogut}}}, \bibinfo {author}
  {\bibfnamefont{S.~S.}\ \bibnamefont{{Meyer}}}, \bibinfo {author}
  {\bibfnamefont{G.~S.}\ \bibnamefont{{Tucker}}}, \bibinfo {author}
  {\bibfnamefont{J.~L.}\ \bibnamefont{{Weiland}}}, \bibinfo {author}
  {\bibfnamefont{E.}~\bibnamefont{{Wollack}}},\ and\ \bibinfo {author}
  {\bibfnamefont{E.~L.}\ \bibnamefont{{Wright}}},\ }%
  \bibfield{journal}{%
  \Doi{10.1088/0067-0049/180/2/330}{\bibinfo {journal} {\apjs}}\ }%
  \textbf{\bibinfo {volume} {180}},\ \bibinfo {pages} {330} (\bibinfo {month}
  {Feb.}\ \bibinfo {year} {2009}),\
  \Eprint{http://arxiv.org/abs/0803.0547}{arXiv:0803.0547}%
  \bibAnnoteFile{NoStop}{2008arXiv0803.0547K}%
\end{thebibliography}%

\end{document}